\documentclass[12pt]{iopart}
\usepackage{hyperref}
\usepackage{lineno}
\usepackage[numbers]{natbib}
\usepackage{notoccite}
\usepackage{caption}
\usepackage{setspace}
\usepackage{geometry}
\usepackage[none]{hyphenat}
\hypersetup{pdfpagemode={UseOutlines},
bookmarksopen=true,
bookmarksopenlevel=0,
hypertexnames=true,
colorlinks=true,
citecolor=blue,
linkcolor=mdtRed,
urlcolor=cyan!90!black!,
pdfstartview={FitV},
unicode,
breaklinks=true,
}
\usepackage{url}
\usepackage{graphicx}
\usepackage{subfigure}

\usepackage{fixme}
\fxsetup{
    status=draft,
    author= ,
    layout=inline,
    theme=color}
\definecolor{fxnote}{rgb}{1,0,0}    
\definecolor{fxwarning}{rgb}{0,1,0} 
\definecolor{fxerror}{rgb}{0,0,1}   

\hypersetup{colorlinks=true,allcolors=blue} 
\begin{document}

\title[Searches for supersymmetry in CMS]{Searches for supersymmetry in CMS}

\author{Uttiya Sarkar for the CMS Collaboration}

\address{Tata Institute of Fundamental Research, Mumbai\\ Homi Bhaba Road, Mumbai 400005, India}
\ead{\href{uttiya.sarkar@cern.ch}{uttiya.sarkar@cern.ch}}
\vspace{11pt}

\begin{indented}
\item March 2021
\end{indented}

\begin{abstract}
The results from the CMS search for supersymmetric particles based on Run-2 data recorded at a center-of-mass energy of 13 TeV are summarized. Strong and weak production of SUSY scenarios are considered. Results presented include the searches for squarks and gluinos, direct production of charginos, neutralinos, and sleptons. These searches involve final state objects comprising jets, missing transverse momentum, electrons or muons, taus or photons, as well as long-lived particles. The data in these searches are found to be consistent with standard model predictions and no significant excess is observed. Upper limits have been set on the masses of supersymmetric particles from a variety of search channels.
\end{abstract}

\vspace{12pc}
\noindent{\it  Ninth International Conference on New Frontiers \\ in Physics (ICNFP 2020)}\\
\noindent{\it  4 Sept.-2 Oct., 2020}\\
%
%
%
%
%

\clearpage

\section{Introduction}
Among the candidate theories beyond the standard model (SM),  supersymmetry (SUSY)~\cite{Fayet:1976cr,Nilles:1983ge,Martin:1997ns} is the most celebrated one thanks to its ability in explaining a wide range of phenomena that are not addressed by the SM. It postulates a symmetry between the fermions and the bosons introducing a set of new particles which differ by spin half-unit from the SM particles. It also posits the existence of the lightest superpartner (LSP) if a new symmetry, namely the R-parity~\cite{Fayet:1974pd}, is invoked. The LSP ($\tilde{\chi}_{1}^{0}$) is considered to be a suitable dark matter candidate~\cite{Farrar:1978xj}. Therefore, energy-frontier experiments such as CMS~\cite{Chatrchyan:2008aa} are focused to venture on experimental signatures of such SUSY theories. Recent results from CMS are exciting with innovative analysis strategies that target a broad spectrum of SUSY production scenarios. In this report, we present a few selected SUSY search results based on the Run-2 data corresponding to a total integrated luminosity of $\sim$ 140 $\mathrm{fb}^{-1}$ collected by the CMS detector.

\section{Searches for Gluino Production}
\begin{figure}
    \centering
    \includegraphics[width=0.9\linewidth]{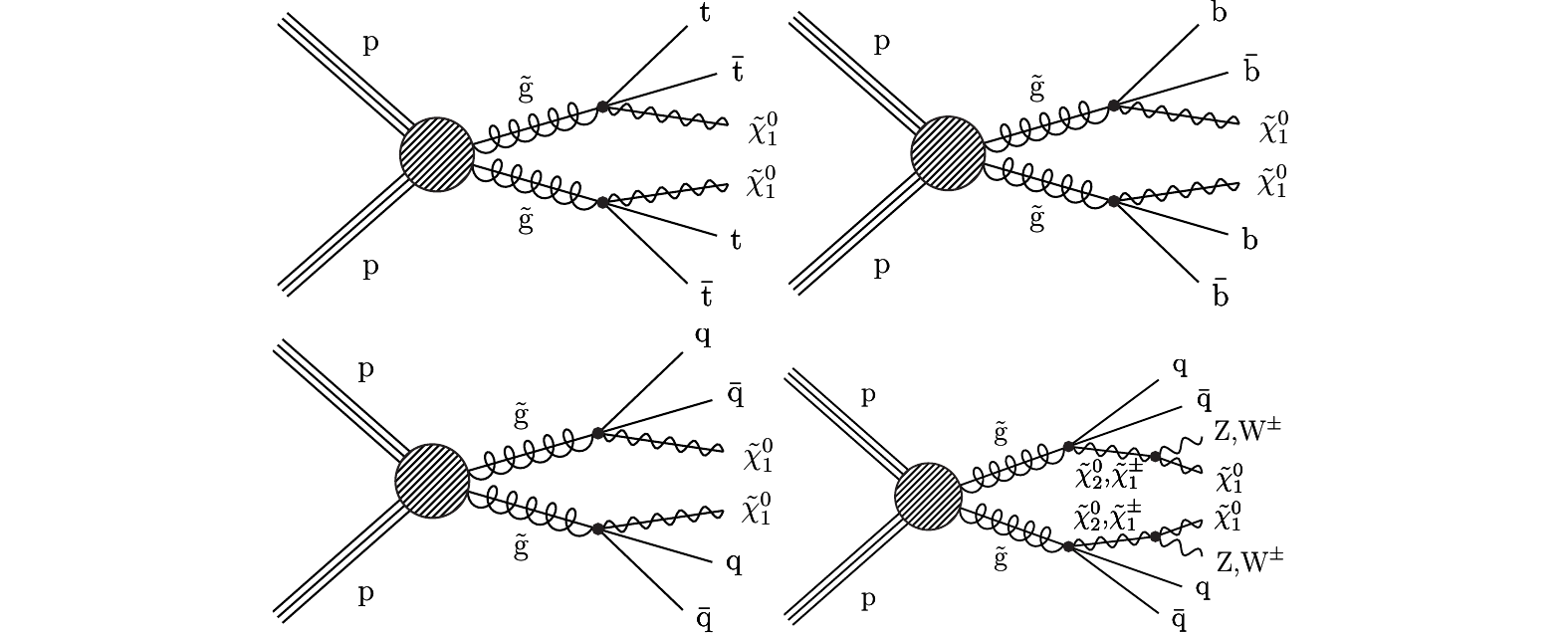}
    \caption{Example event diagrams of the gluino pair production at the LHC, where the gluinos decay into heavy quarks and $\tilde{\chi}^0_1$ (top), light quarks (bottom left) and $\tilde{\chi}^0_1$, and light quarks and SM vector bosons (bottom right).}
    \label{fig:001}
\end{figure}

Strong production of SUSY particles can be searched for at higher masses due to their larger cross sections as compared to the other production modes. In particular, gluinos ($\tilde{g}$) -- superpartners of gluons -- are produced with the highest cross section at the LHC. Their searches are typically characterized by large jet multiplicity $\mathrm{N_{jets}}$ and the presence of large missing transverse momentum ($\mathrm{p_T^{miss}}$).  Additional discriminating variables are the number of leptons or photons and the b-tagged jets $\mathrm{N_{b-jets}}$ in the final state. The other kinematic variables that are also exploited are the scalar sum of jet transverse momenta ($\mathrm{H_T}$), jet masses, variables related to large-radius jets, etc. A few example event diagrams of the gluino pair production at the LHC are shown in Fig.~\ref{fig:001}.

\begin{figure}
    \centering
    \includegraphics[width=0.4\linewidth]{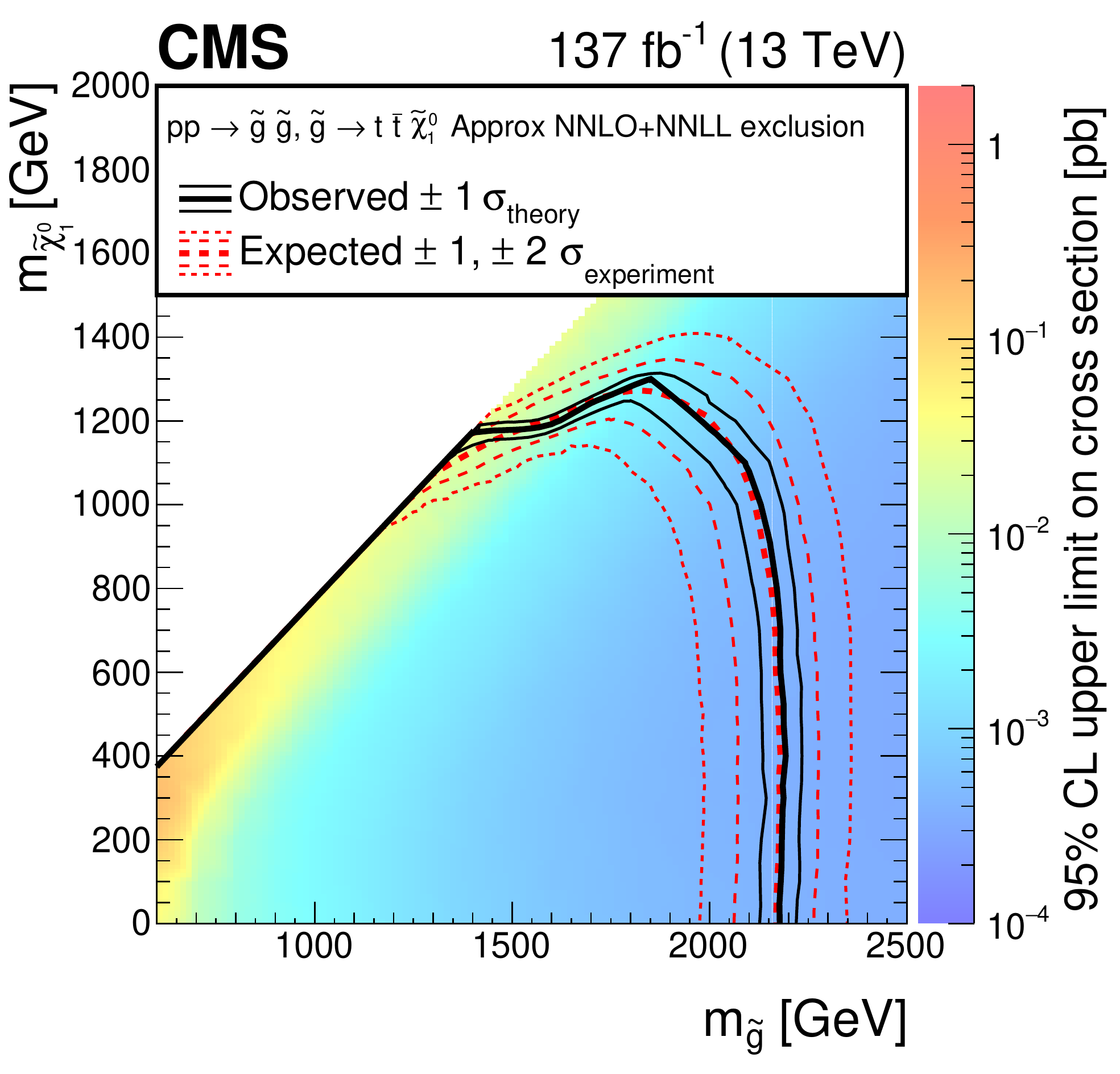}
    \includegraphics[width=0.4\linewidth]{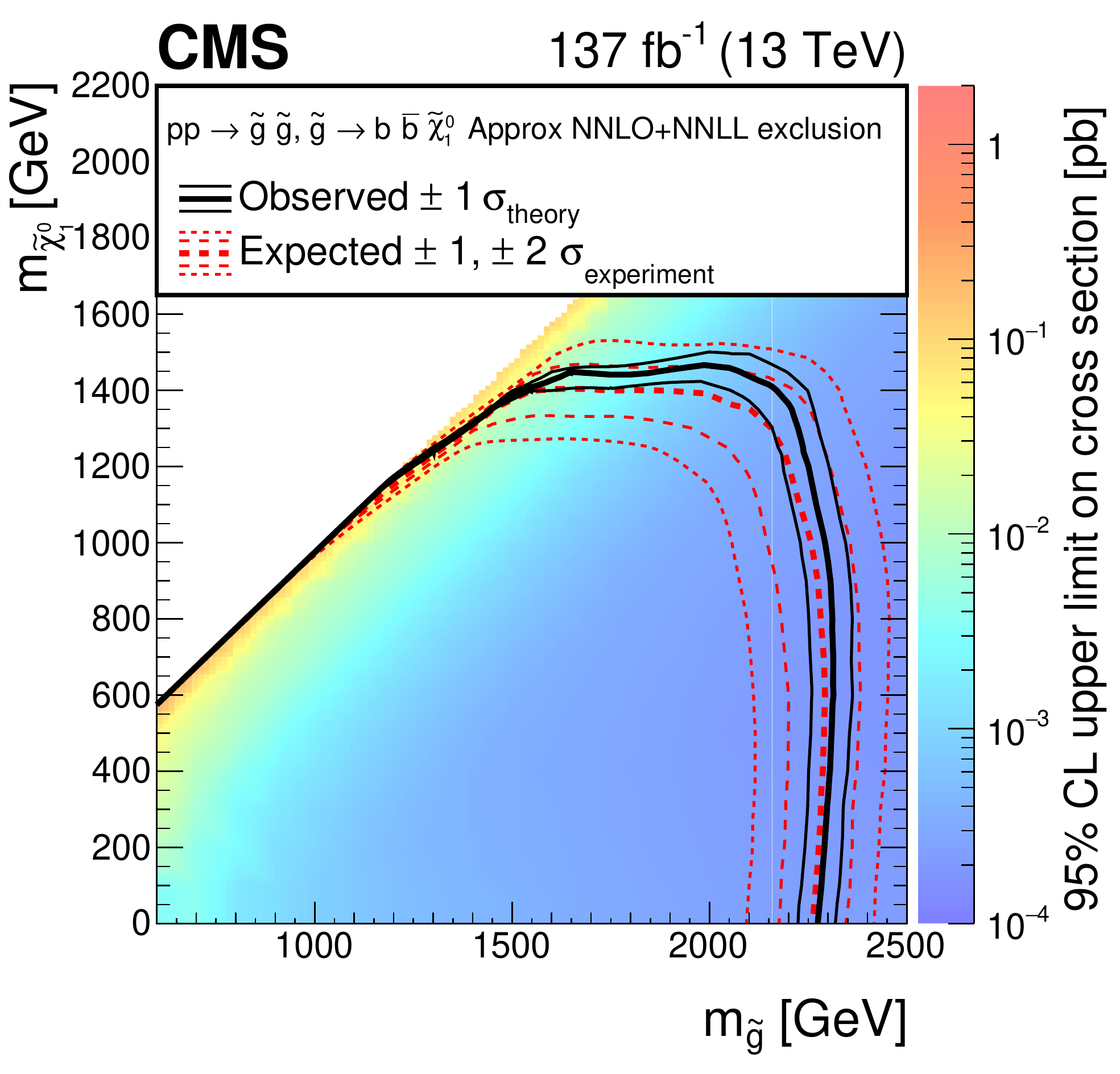}
    \caption{Exclusion limits at 95\% CL on gluino masses at different neutralino ($\tilde{\chi}^0_1$) mass points for gluino to top (left) and bottom (right) quark decay modes from inclusive hadronic searches based on the jet multiplicity and missing transverse momentum~\cite{Sirunyan:2019ctn}. The solid black curve represents the observed exclusion and the dashed red curve represents the expected exclusion limits in the plane of gluino and $\tilde{\chi}^0_1$ masses.}
    \label{fig:002}
\end{figure}

Searches pursued using complementary analysis strategies provide independent exclusion limits on the gluino production cross section. A classic example~\cite{Sirunyan:2019ctn} involves pair-produced gluinos decaying into purely hadronic final states. The aforementioned variables, namely $\mathrm{N_{jets}}$, $\mathrm{N_{b-jets}}$, $\mathrm{H_T}$ and $\mathrm{p_T^{miss}}$ are used to divide the available search region into 174 independent bins. The search excludes gluino masses up to 2310 GeV at 95\% confidence level (CL) as shown in Fig.~\ref{fig:002}.

\begin{figure}
    \centering
    \includegraphics[width=0.5\linewidth]{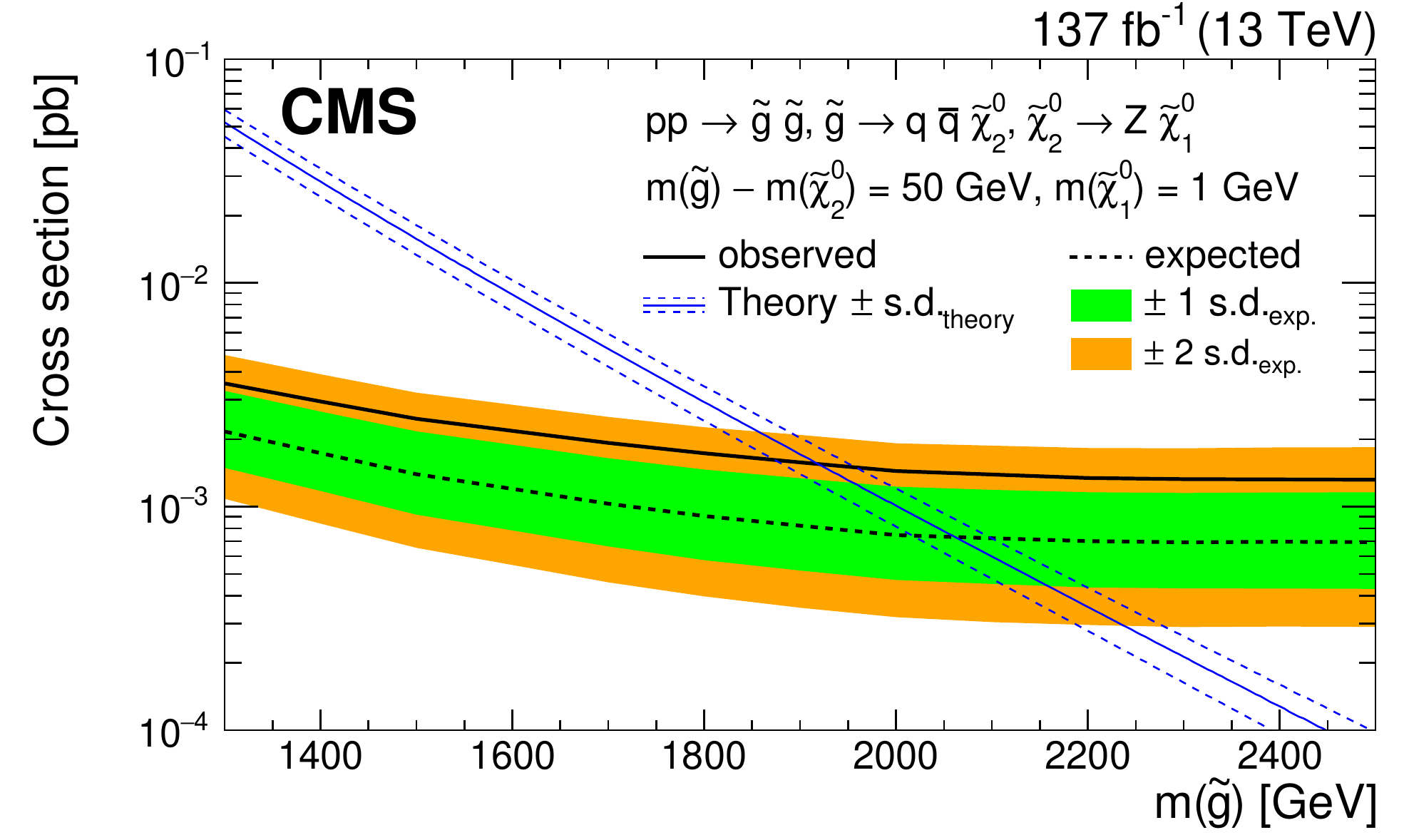}
    \caption{Exclusion limits at 95\% CL on gluino mass from the events with high-momentum Z bosons and missing transverse momentum~\cite{Sirunyan:2020zzv}. The solid black curve shows the observed exclusion limit.  The dashed black curve shows the expected limit and the green and yellow bands represent the $\pm$1 and $\pm$2 standard deviation uncertainty ranges. }
    \label{fig:003}
\end{figure}

Another interesting search~\cite{Sirunyan:2020zzv} exploits the topology of large-radius jets in the final state that can provide a common decay scenario for a TeV scale ($\simeq$ 3 TeV) initial SUSY particle. In this search, a pair of gluinos is produced, each decaying into a high momentum ($\mathrm{p_T}>200$ GeV) Z boson and $\mathrm{p_T^{miss}}$. A small mass splitting between the gluino and the intermediate neutralino ensures a large momentum for the final state Z bosons, provided that the LSP mass is fixed at 1 GeV. The Z boson gives rise to a large radius jet providing a narrow search window in the jet mass. The latter variable is fitted with a linear function to predict the number of background events in the search region. An exclusion limit on the gluino mass of 1920 GeV (Fig.~\ref{fig:003}) is achieved with this novel search strategy.

\begin{figure}
    \centering
    \includegraphics[width=0.4\linewidth]{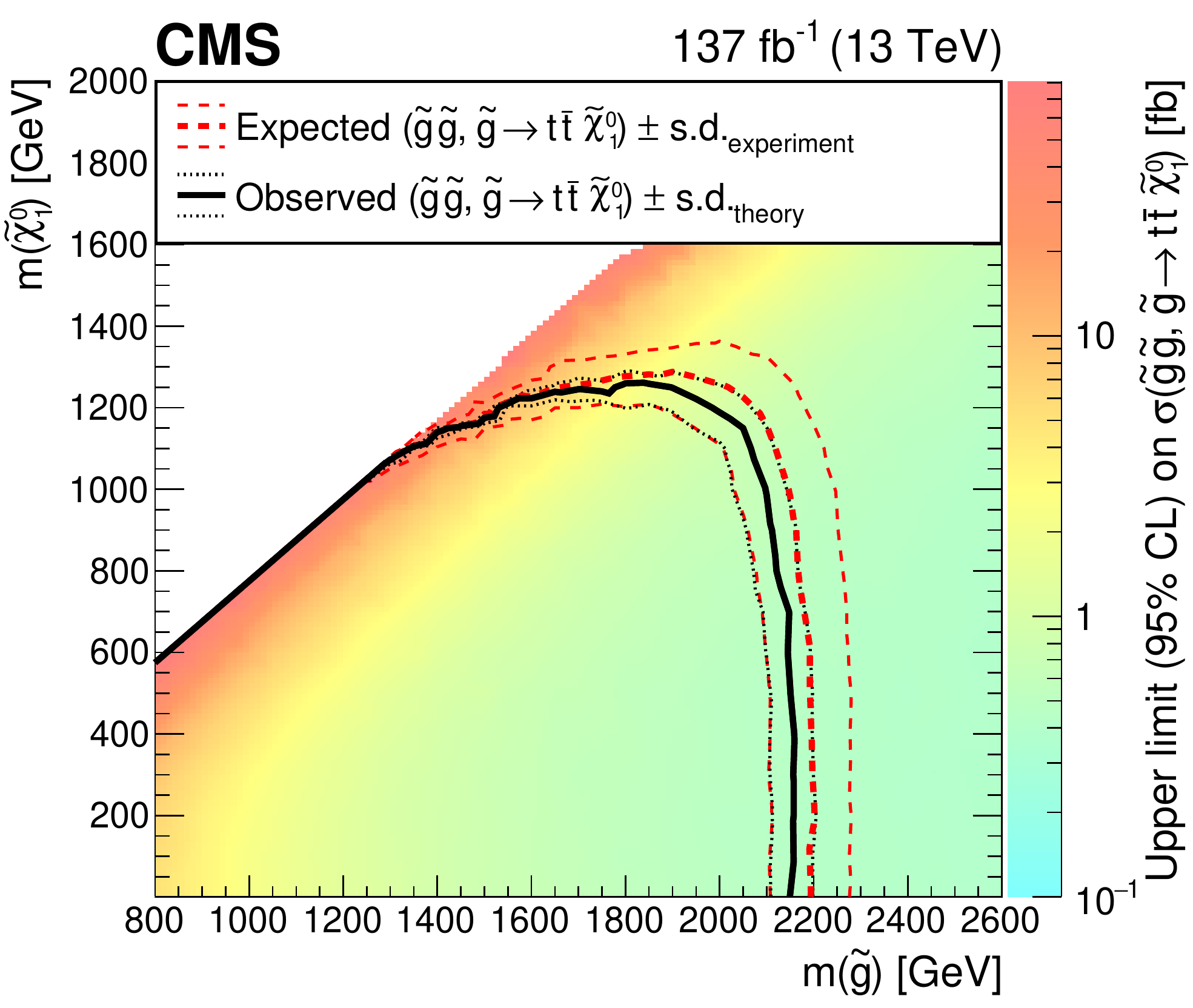}
    \caption{Exclusion limits at 95\% CL on gluino masses from the SUSY search in the final states with a single lepton using the sum of masses of large-radius jets~\cite{CMS:2019tlp}.}
    \label{fig:004}
\end{figure}

Unlike the previous ones described, the analysis that deals with one lepton and the sum of masses of large-radius jets final states~\cite{CMS:2019tlp}, makes use of the jets emanating from initial-state radiation. The sum of the masses of large-radius jets ($\mathrm{M_{jet}}$) and the transverse mass ($\mathrm{m_T}$) variables are used to categorize the search and control regions. An ABCD method is deployed in these two variables to estimate the background events in the search regions. Scenarios with gluino masses up to about 2150 GeV are excluded at 95\% CL as shown in Fig.~\ref{fig:004}. A similar level of exclusion on the gluino mass $\sim$2 TeV is achieved by other search strategies~\cite{Sirunyan:2020ztc,Sirunyan:2019xwh}.

\section{Searches for Stop Production}
\begin{figure}
    \centering
    \includegraphics[width=0.8\linewidth]{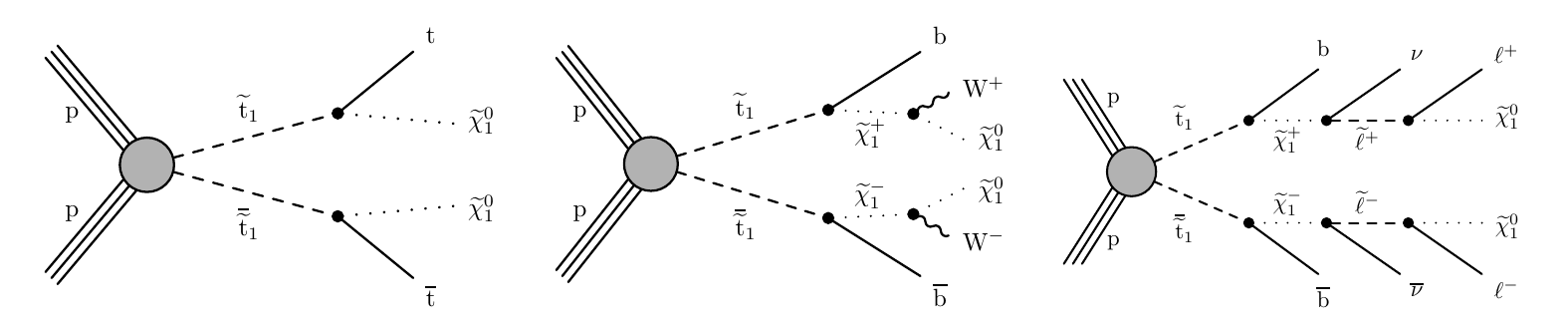}
    \caption{Example event diagrams of the stop pair production at the LHC, where the stop decays into a SM top and $\tilde{\chi}^0_1$ (left), SM bottom, W boson and $\tilde{\chi}^0_1$ (middle), and SM bottom, neutrinos, leptons and $\tilde{\chi}^0_1$ (right).}
    \label{fig:005}
\end{figure}

It is important for a SUSY theory to be ``natural", which means that there are minimal fine-tunings of the parameters of the theory. A ``natural SUSY"~\cite{Moreno:2015zeg} scenario introduces mildest possible fine-tuning. The searches for the third generation stop ($\tilde{t}$) production are of special importance from such naturalness arguments~\cite{Barger:2012hr,KING2012207}.  A few example event diagrams for different decay modes are shown in Fig.~\ref{fig:005}.

\begin{figure}
    \centering
    \includegraphics[width=0.8\linewidth]{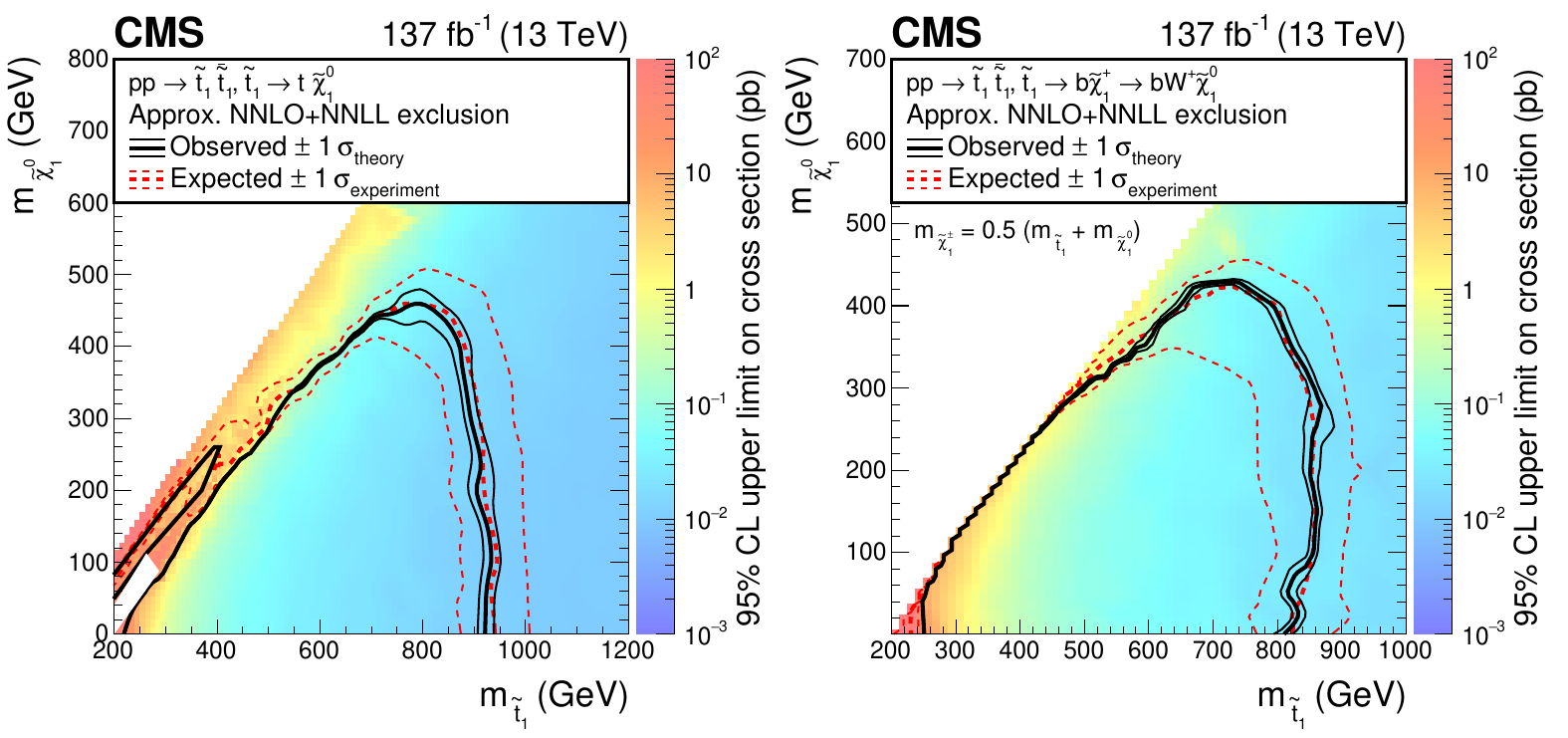}
    \caption{Exclusion limits at 95\% CL on top squark masses from the SUSY search in the final states with two opposite charge leptons~\cite{Sirunyan:2020tyy}. The left plot shows the limit for top squark decaying into top and $\tilde{\chi}^0_1$, whereas the right plot shows the top squark decaying into a W boson and $\tilde{\chi}^0_1$ via an intermediate chargino production. The solid black curve shows the observed exclusion and the red dashed curve represents the expected exclusion on the plane of $\tilde{\chi}^0_1$ and top squark masses. }
    \label{fig:006}
\end{figure}

The search for top squark pair production using dilepton final states~\cite{Sirunyan:2020tyy} focuses on the requirement of the two opposite charged leptons ($\mu$,e) in the final state. Further, the search region is binned in transverse mass ($\mathrm{M_{T_2}}$) and $\mathrm{p_{T}^{miss}}$ significance (S) variables. A lower limit at 95\% CL on the top squark mass is placed up to 925 GeV as shown in Fig.~\ref{fig:006}.

\begin{figure}
    \centering
    \includegraphics[width=0.4\linewidth]{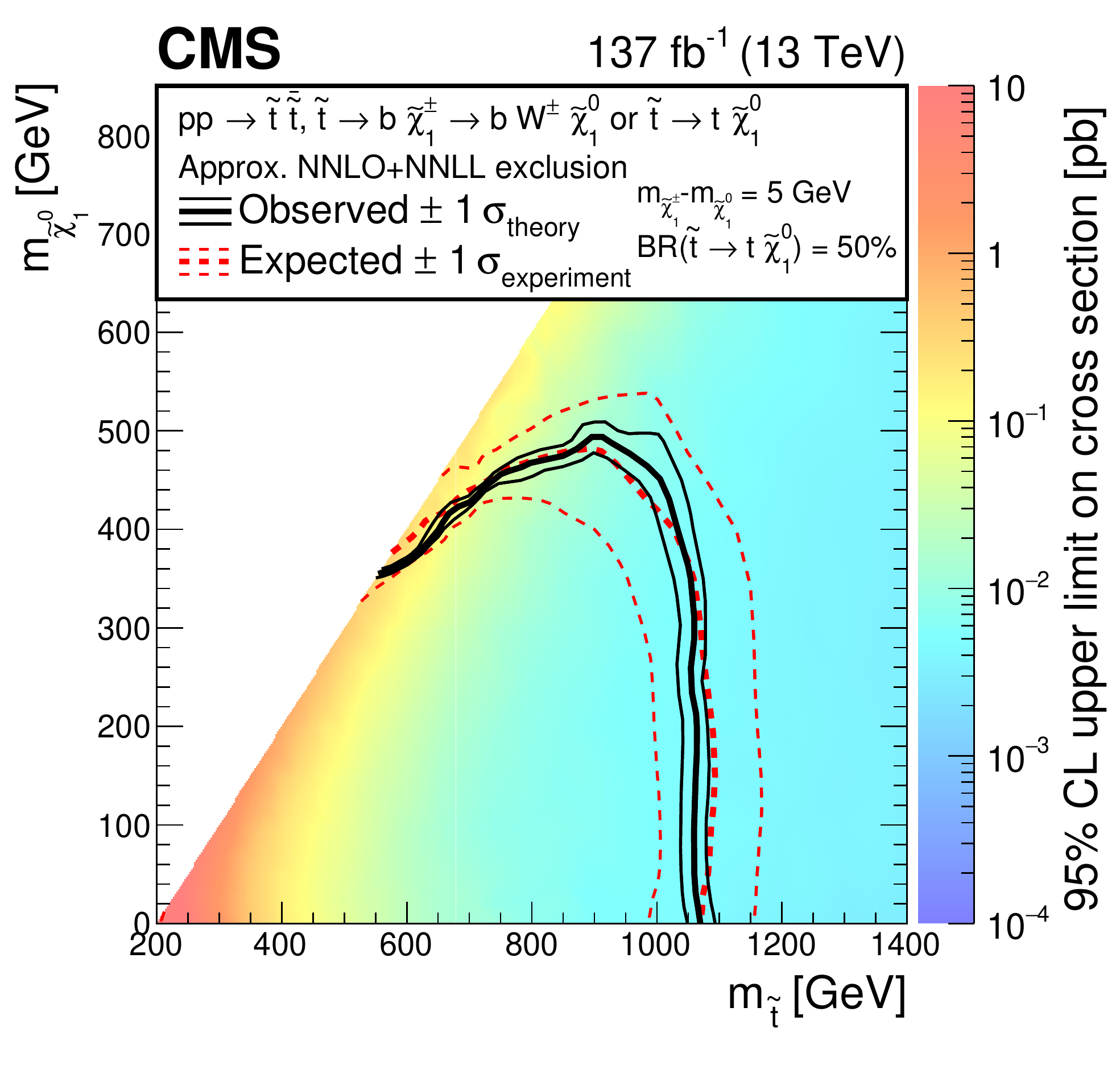}
    \caption{Exclusion limits at 95\% CL on top squark masses from the SUSY search with one lepton, jets, and missing transverse momentum~\cite{Sirunyan:2019glc}, where the top squark decays into either top and $\tilde{\chi}^0_1$ or a bottom quark,  boson and $\tilde{\chi}^0_1$ via an intermediate chargino. The mass splitting between the chargino and the $\tilde{\chi}^0_1$ is set at 5 GeV and the branching ratios of each of the decay scenarios is 50\%. }
    \label{fig:007}
\end{figure}

Another search for direct top squark pair production involves events with one lepton, multiple jets, and large transverse momentum imbalance~\cite{Sirunyan:2019glc}. An exclusion limit at 95\% CL is set on cross sections for the top squark masses at 1.2 TeV as shown in Fig.~\ref{fig:007}. A similar level of exclusion on the stop mass $\sim$1 TeV is achieved by other search strategies~\cite{Sirunyan:2019ctn,Sirunyan:2020ztc,Sirunyan:2019xwh}.

\section{Searches for Electroweak Production}
\begin{figure}[htbp]
    \centering
    \includegraphics[width=0.5\linewidth]{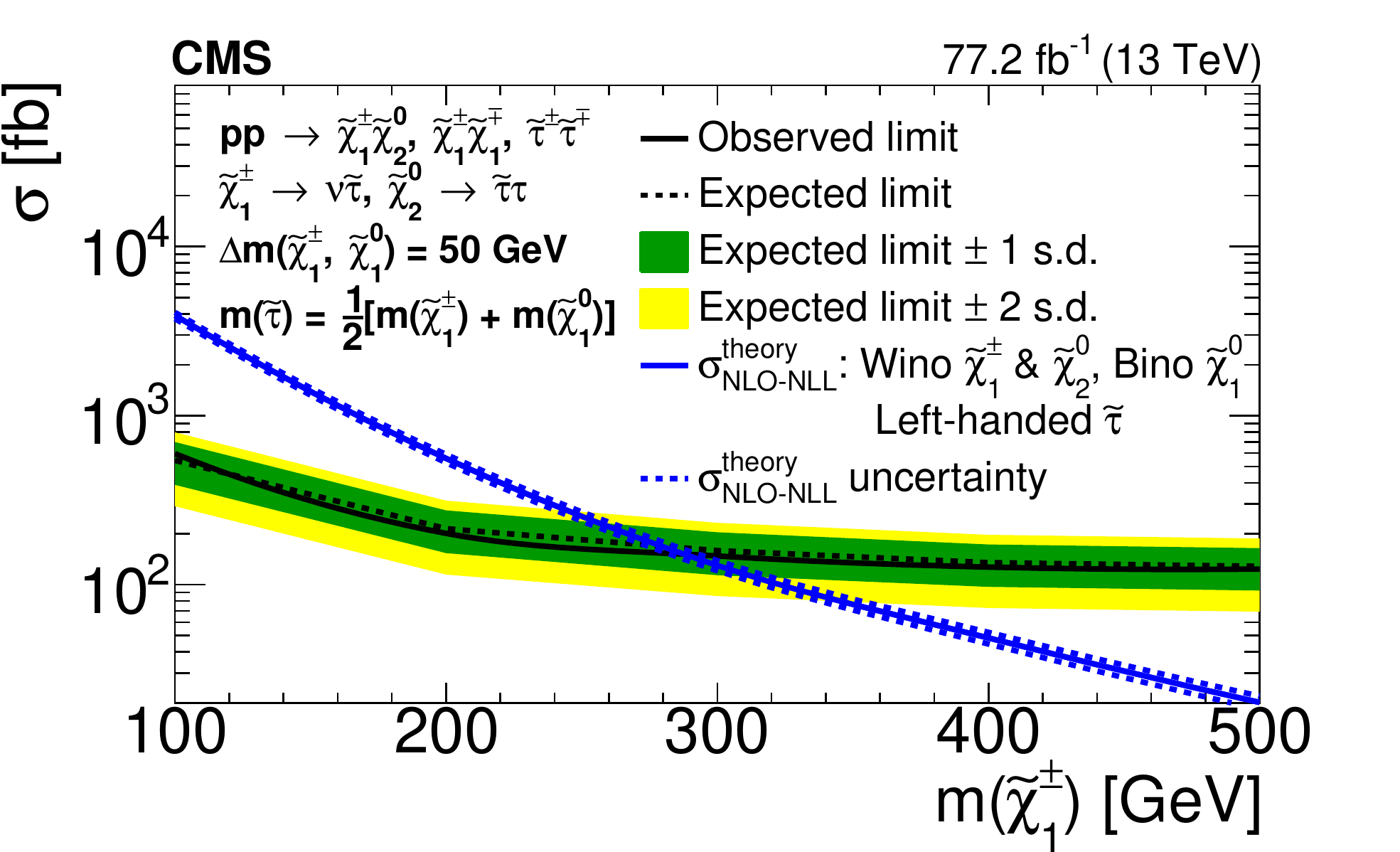}
    \caption{Exclusion limits at 95\% CL on the chargino production cross section in events with a soft $\tau$ lepton, a highly energetic jet, and large missing transverse momentum~\cite{Sirunyan:2019mlu}. The blue curve represents the theoretical cross sections, whereas the solid black curve shows the observed exclusion limit. The dashed black curve shows the expected exclusion limit and the green and yellow bands represent the $\pm$1 and $\pm$2 standard deviation on the prediction, respectively.}
    \label{fig:008}
\end{figure}

The present LHC luminosity limits probing electroweak decay scenarios of SUSY particles due to their low production cross sections. Despite this, several interesting channels exist that provide very clean signatures. Many scenarios have been looked at so far, including the direct production of chargino/neutralino, gauge-mediated, or direct production of weakly interacting massive particles. The interesting models are R-parity conserving ones where the LSP could undergo annihilation-production interactions with SM particles and providing the observed dark matter relic density, assuming coannihilation models of the stau ($\tilde{\tau}$) and the $\tilde{\chi}^0_1$~\cite{Spergel:2003cb,planck}. One such search is based on a compressed mass spectrum in events with a soft $\tau$ lepton, a highly energetic jet, and large missing transverse momentum~\cite{Sirunyan:2019mlu}. The results are interpreted with a small mass difference (50 GeV) between the chargino and the neutralino. The mass of the top squark is assumed to be the average of the chargino and neutralino masses. The data show no deviation from the SM and upper limits at 95\% CL are set on the sum of the chargino, neutralino, and the stop production cross sections, resulting in a lower limit of 290 GeV on the mass of the chargino. This is the most stringent limit available to the date surpassing the bounds from LEP experiments. This exclusion limit is shown in Fig.~\ref{fig:008}.

\section{Searches for Long Lived Particles}
\begin{figure}[htpb]
    \centering
    \includegraphics[width=0.4\linewidth]{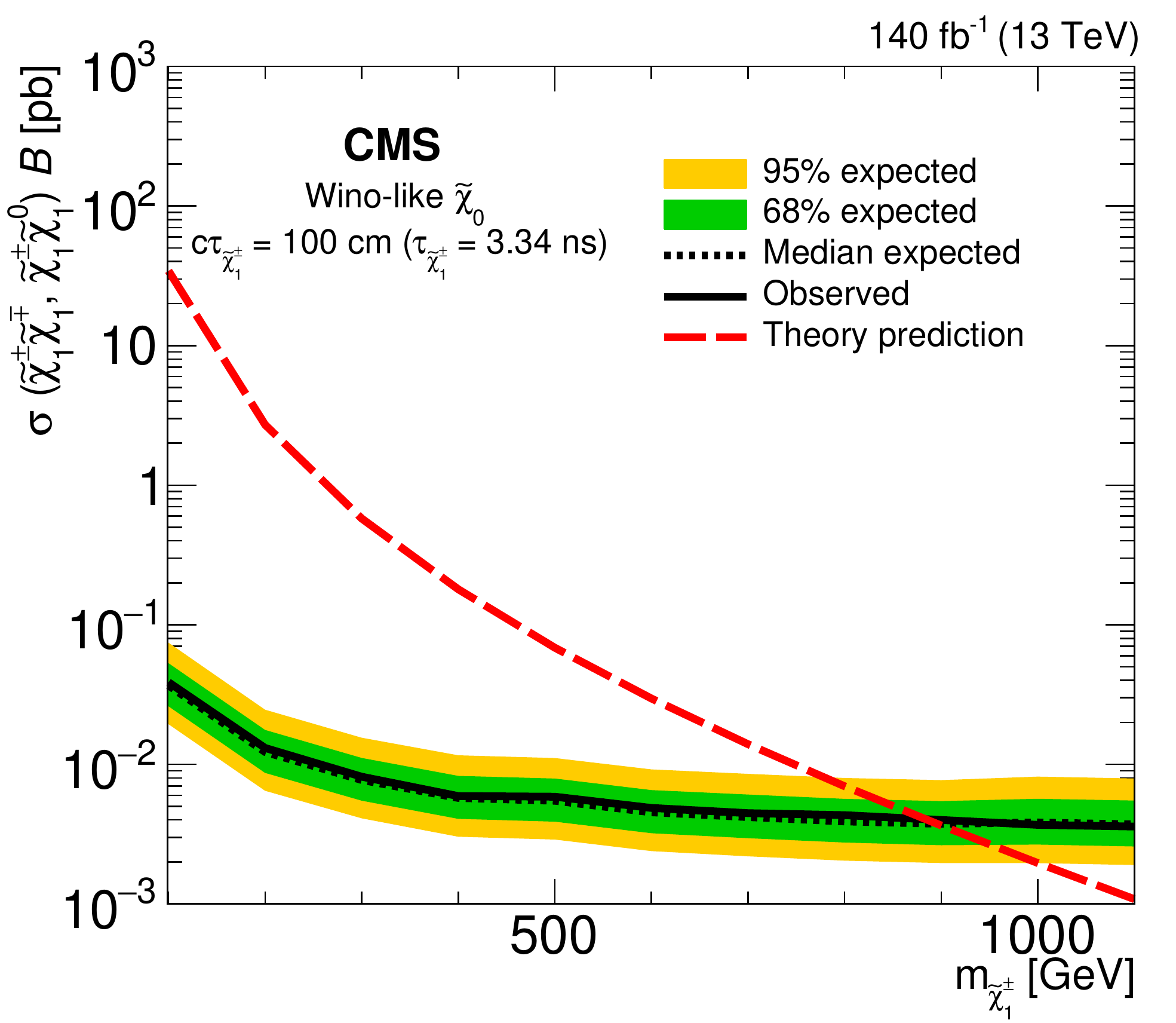}
    \includegraphics[width=0.4\linewidth]{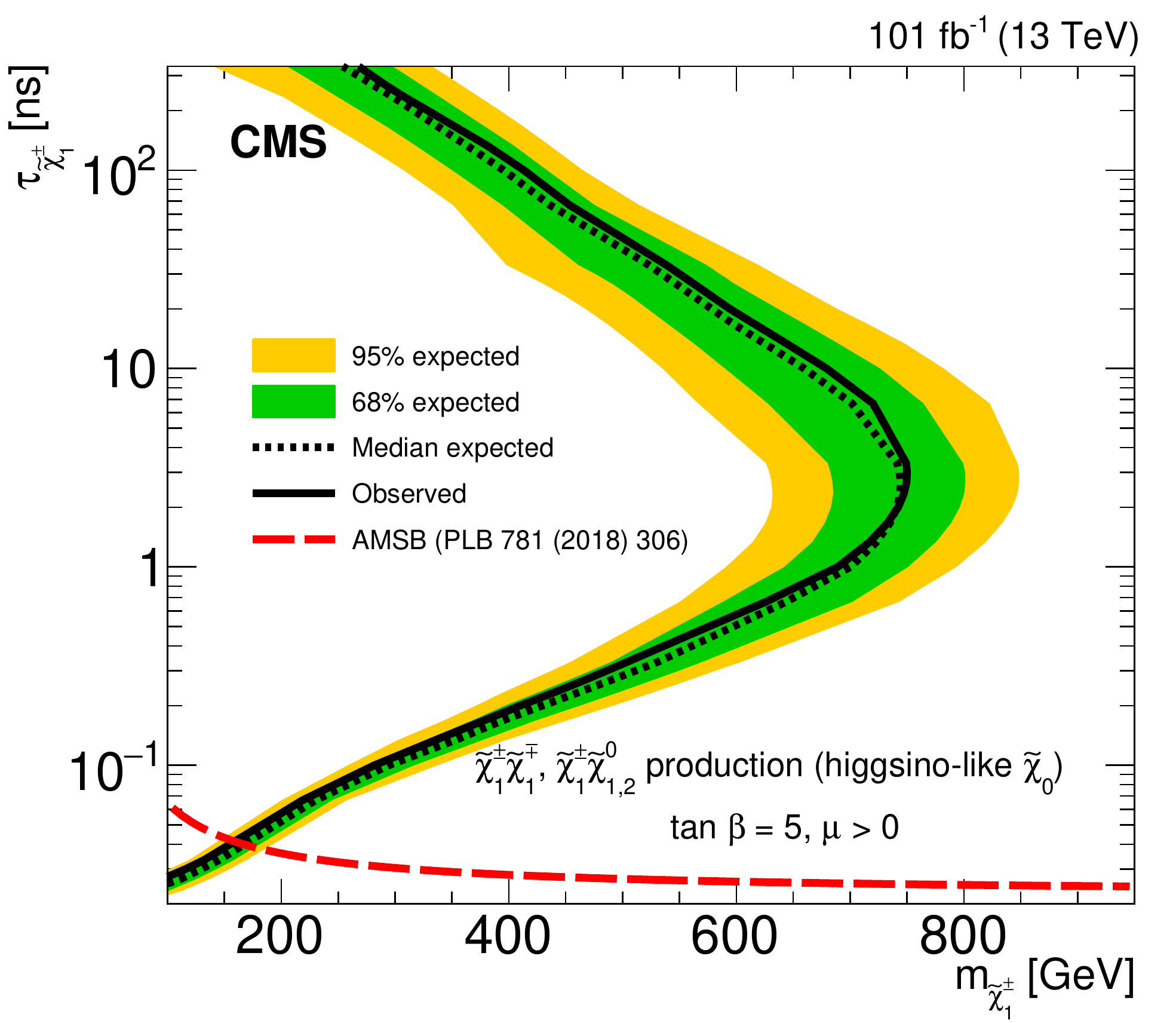}
    \caption{Exclusion limits at 95\% CL on chargino masses from the wino-like production scenario for a chargino lifetime of 3.34 ns (left) and higgsino-like  production scenario (right) in searches with disappearing tracks ~\cite{Sirunyan:2020pjd}. }
    \label{fig:009}
\end{figure}

A whole enterprise of long-lived particle (LLP) searches is becoming more popular across the collider community since LLPs are a common signature for a wide range of beyond-the-SM theories. Such searches are usually signature-driven that give rise to disappearing tracks or displaced vertices. One such exciting result for the Run-2 data is from the search for disappearing tracks~\cite{Sirunyan:2020pjd}.
The upper limits are set on chargino production in the context of an anomaly-mediated SUSY breaking model for purely wino and higgsino neutralino scenarios. The chargino mass is excluded at 95\% CL up to  884 (474) GeV for a lifetime of 3 (0.2) ns for a wino case and below 750 (175) GeV for a lifetime of 3 (0.05) ns for a higgsino case as shown in Fig.~\ref{fig:009}.
\section{Summary and Future Prospects}
\begin{figure}[h]
    \centering
    \includegraphics[width=0.32\linewidth]{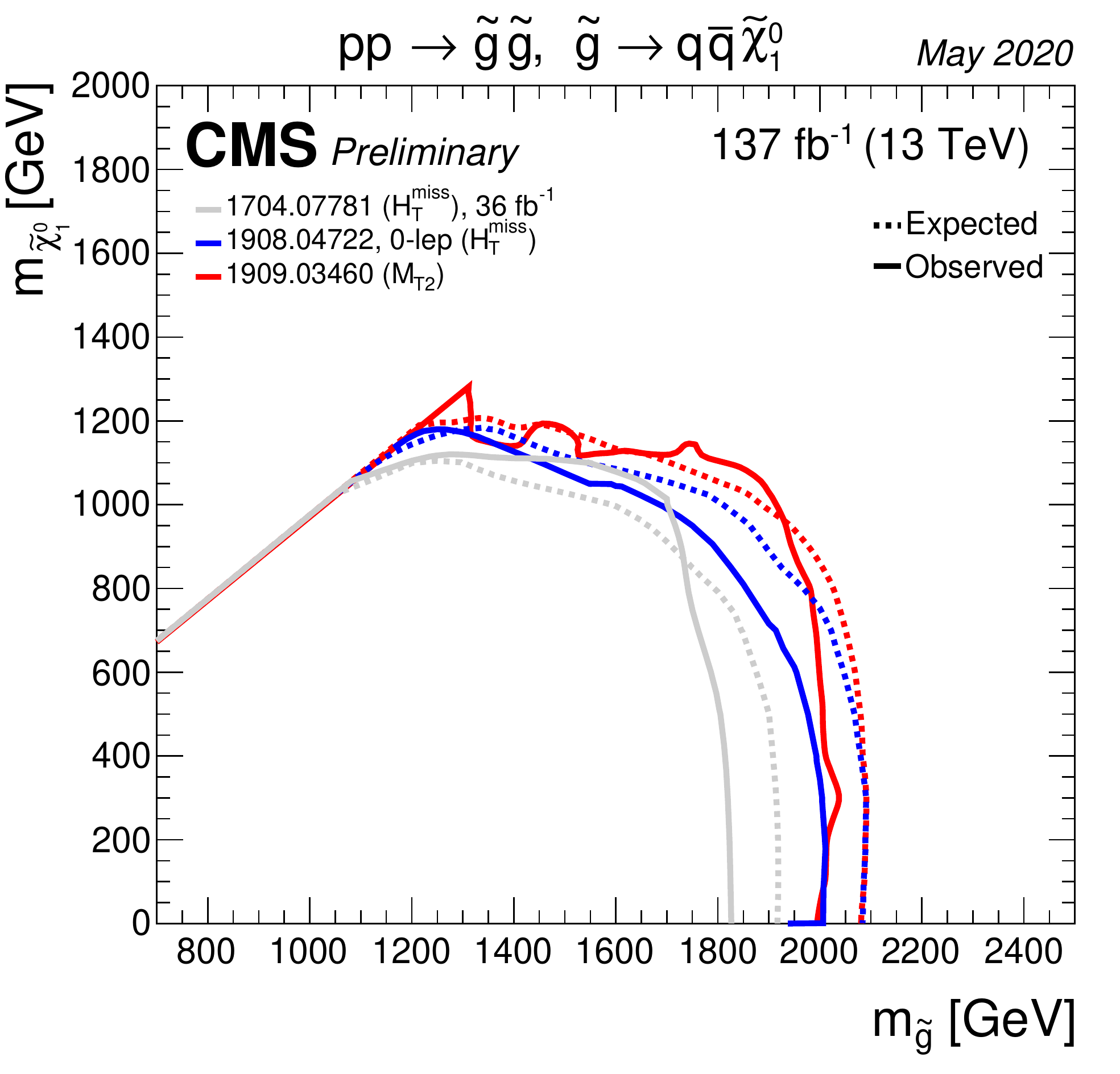}
    \includegraphics[width=0.32\linewidth]{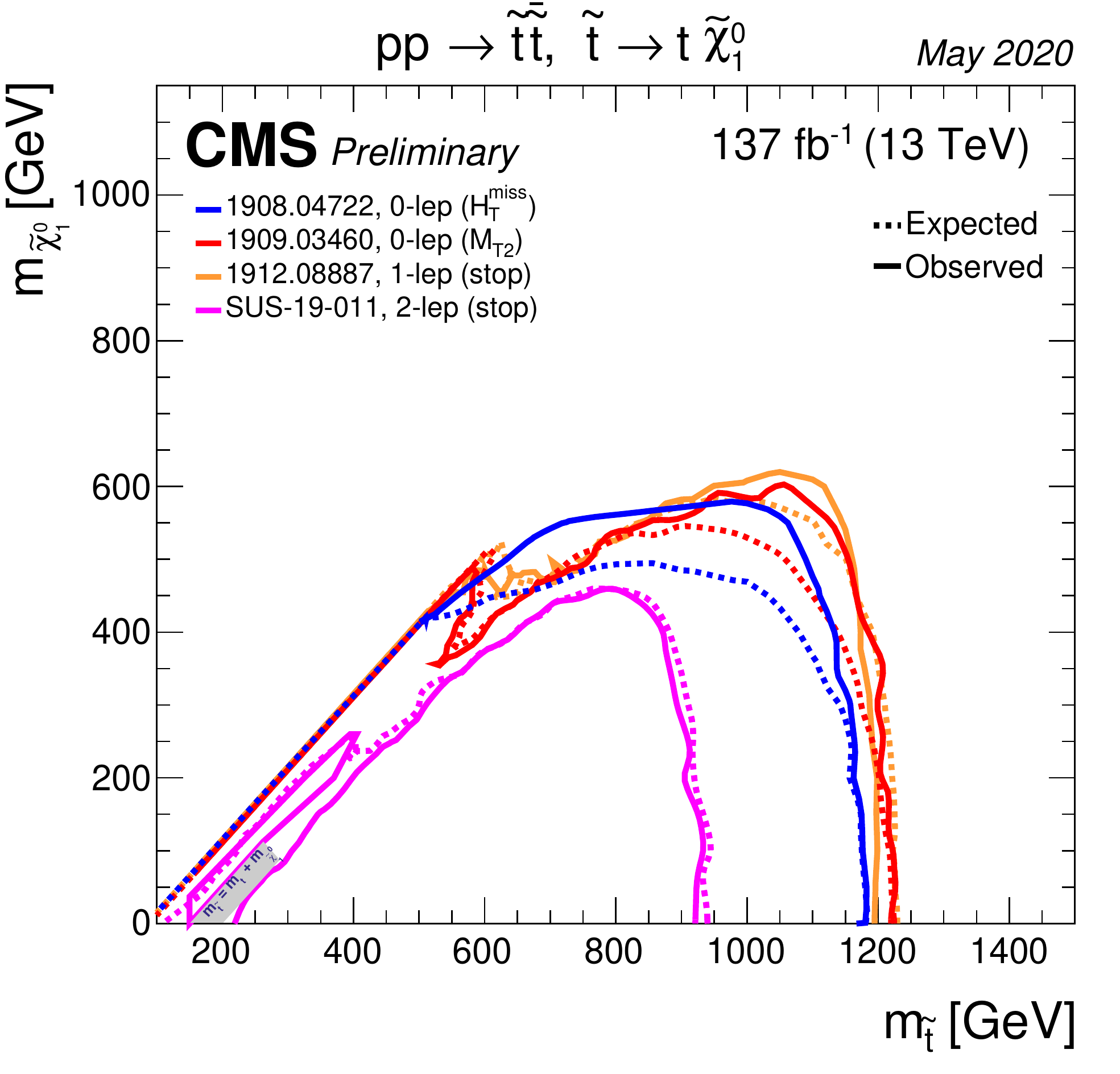}
    \includegraphics[width=0.32\linewidth]{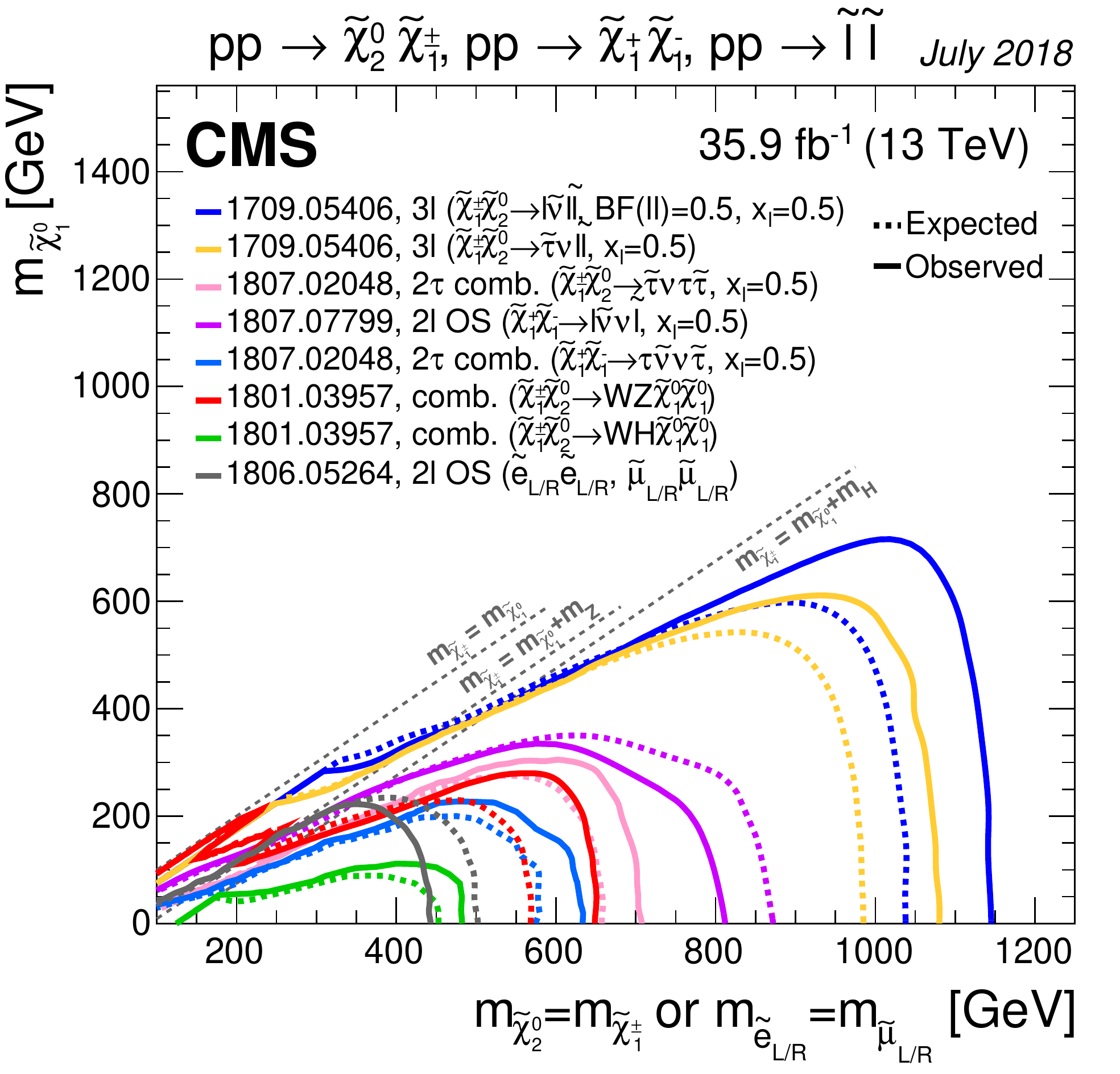}
    \caption{Summary of limits obtained by CMS for pair produced gluinos (left), pair produced top squark (middle) and electroweak searches (right)~\cite{Suspubliccms}.}
    \label{fig:011}
\end{figure}

With a delivered luminosity of $\sim$140 $\mathrm{fb}^{-1}$ in Run-2, CMS has provided very stringent limits on several SUSY production scenarios. So far, no significant deviation from the SM is observed. The exclusion limits on the gluino masses are set at almost 2 TeV while that on the top squark masses are as large as 1.2 TeV. Stronger limits on the electroweak searches for slepton mediated decays exclude charginos between 0.6-1.1 TeV, followed by the boson mediated ones that exclude charginos of 0.4-0.7 TeV, whereas weaker limits for compressed regions and direct stau pair production are set at 0.3 TeV. A summary of the CMS results is presented in Fig.~\ref{fig:011}.

With the upgraded High-Luminosity LHC scenario (HL-LHC) at Run-4 we expect more exciting results from CMS. An expected exclusion reach from HL-LHC on gluino masses is about 3 TeV and on stop masses at 2 TeV~\cite{yellowreport}. With the expected increment in luminosity, the electroweak and LLP searches would also become even more promising.

\bibliographystyle{unsrtnat}
\bibliography{refs}

\end{document}